\documentclass[aps,prl,twocolumn,showpacs,superscriptaddress,preprintnumbers]{revtex4}  
\usepackage{graphicx}  
\usepackage{dcolumn}   
\usepackage{bm}        
\usepackage{amssymb}   
\usepackage{amsmath}   
\usepackage{multirow}
\usepackage{color}
\usepackage{units}
\usepackage{xspace}

\newcommand{\deltacp}{\ensuremath{\delta_{\rm CP}}}

\newcommand{\lidresult}{6}
\newcommand{\lidbgpredict}{\ensuremath{0.99\pm0.11{\rm~(syst.)}}}
\newcommand{\lidsigmaapp}{\ensuremath{3.3\sigma}}
\newcommand{\liddcprange}{\ensuremath{0.1\pi<\deltacp{}<0.5\pi}}

\newcommand{\lemresult}{11}
\newcommand{\lembgpredict}{\ensuremath{1.07\pm0.14{\rm~(syst.)}}}
\newcommand{\lemsigmaapp}{\ensuremath{5.3\sigma}}
\newcommand{\lemdcprangenh}{\ensuremath{0.25\pi<\deltacp{}<0.95\pi}}

\newcommand{\etal}{{\it et al.}\xspace}

\newcommand{\nova}{NOvA}
\newcommand{\numu}{\ensuremath{\nu_{\mu}}\xspace}
\newcommand{\nue}{\ensuremath{\nu_{e}}\xspace}
\newcommand{\numubar}{\ensuremath{\bar{\nu}_{\mu}}\xspace}
\newcommand{\nuebar}{\ensuremath{\bar{\nu}_{e}}\xspace}
\newcommand{\numutonue}{\ensuremath{\nu_\mu \rightarrow \nu_e}\xspace}
\newcommand{\nueCC}{\ensuremath{\nu_e\,{\rm CC}}\xspace}
\newcommand{\numuCC}{\ensuremath{\nu_{\mu}\,{\rm CC}}\xspace}
\newcommand{\nutau}{\ensuremath{\nu_{\tau}}\xspace}
\newcommand{\dmsq}[1]{\ensuremath{\Delta m^2_{ #1 }}\xspace}
\newcommand{\sinsq}[1]{\ensuremath{\sin^{2}2\theta_{ #1 }}\xspace}
\newcommand{\sinsqnotwo}[1]{\ensuremath{\sin^{2}\theta_{ #1 }}\xspace}

\newcommand{\piz}{\mbox{$\pi^{0}$}}      
\newcommand{\nutauCC}{\ensuremath{\nu_{\tau}\,{\rm CC}}\xspace}      
\newcommand{\POT}{POT}
\newcommand{\kton}{kt}

\newcommand{\nh}{NH}
\newcommand{\ih}{IH}

\begin{document}

\title{First measurement of electron neutrino appearance in NOvA}

\newcommand{\ANL}{Argonne National Laboratory, Argonne, Illinois 60439, 
USA}
\newcommand{\IOP}{Institute of Physics, The Czech Academy of Sciences, 
Prague, Czech Republic}
\newcommand{\Athens}{Department of Physics, University of Athens, 
Athens, 15771, Greece}
\newcommand{\BHU}{Department of Physics, Banaras 
Hindu University, Varanasi, 221 005, India}
\newcommand{\UCLA}{Physics and Astronomy Department, UCLA, Box 951547, Los 
Angeles, California 90095-1547, USA}
\newcommand{\Caltech}{California Institute of 
Technology, Pasadena, California 91125, USA}
\newcommand{\Cochin}{Department of Physics, Cochin University
of Science and Technology, Kochi 682 022, India}
\newcommand{\Charles}
{Charles University in Prague, Faculty of Mathematics and Physics,
 Institute of Particle and Nuclear Physics, Prague, Czech Republic}
\newcommand{\Cincinnati}{Department of Physics, University of Cincinnati, 
Cincinnati, Ohio 45221, USA}
\newcommand{\CTU}{Czech Technical University in Prague,
Brehova 7, 115 19 Prague 1, Czech Republic}
\newcommand{\Dallas}{Physics Department, University of Texas at Dallas,
800 W. Campbell Rd. Richardson, Texas 75083-0688, USA}
\newcommand{\Delhi}{Department of Physics \& Astrophysics, University of 
Delhi, Delhi 110007, India}
\newcommand{\JINR}{Joint Institute for Nuclear Research Joliot-Curie, 6 
Dubna, Moscow region 141980, Russia}
\newcommand{\FNAL}{Fermi National Accelerator Laboratory, Batavia, 
Illinois 60510, USA}
\newcommand{\UFG}{Instituto de F\'{i}sica, Universidade Federal de 
Goi\'{a}s, Goi\^{a}nia, GO, 74690-900, Brazil}
\newcommand{\Guwahati}{Department of Physics, IIT Guwahati, Guwahati, 781 
039, India}
\newcommand{\Harvard}{Department of Physics, Harvard University, 
Cambridge, Massachusetts 02138, USA}
\newcommand{\IHyderabad}{Department of Physics, IIT Hyderabad, Hyderabad, 
502 205, India}
\newcommand{\Hyderabad}{School of Physics, University of Hyderabad, 
Hyderabad, 500 046, India}
\newcommand{\Indiana}{Indiana University, Bloomington, Indiana 47405, 
USA}
\newcommand{\INR}{Inst. for Nuclear Research of Russian, Academy of 
Sciences 7a, 60th October Anniversary prospect, Moscow 117312, Russia}
\newcommand{\Iowa}{Department of Physics and Astronomy, Iowa State 
University, Ames, Iowa 50011, USA}
\newcommand{\Jammu}{Department of Physics and Electronics, University of 
Jammu, Jammu Tawi, 180 006, J\&K, India}
\newcommand{\Lebedev}{Nuclear Physics Department, Lebedev Physical 
Institute, Leninsky Prospect 53, 119991 Moscow, Russia}
\newcommand{\MSU}{Department of Physics \& Astronomy, Michigan State 
University, East Lansing, Michigan 48824, USA}
\newcommand{\Duluth}{Department of Physics \& Astronomy, 
University of Minnesota - Duluth, Duluth, Minnesota 55812, USA}
\newcommand{\Minnesota}{School of Physics and Astronomy, University of 
Minnesota - Twin Cities, Minneapolis, Minnesota 55455, USA}
\newcommand{\Oxford}{Subdepartment of Particle Physics, 
University of Oxford, Oxford OX1 3RH, United Kingdom}
\newcommand{\Panjab}{Department of Physics, Panjab University, 
Chandigarh, 106 014, India}
\newcommand{\RAL}{Rutherford Appleton Laboratory, Science and 
Technology Facilities Council, Didcot, OX11 0QX, United Kingdom}
\newcommand{\Carolina}{Department of Physics and Astronomy, University of 
South Carolina, Columbia, South Carolina 29208, USA}
\newcommand{\SDakota}{South Dakota School of Mines and Technology, Rapid 
City, South Dakota 57701, USA}
\newcommand{\SMU}{Department of Physics, Southern Methodist University, 
Dallas, Texas 75275, USA}
\newcommand{\Stanford}{Department of Physics, Stanford University, 
Stanford, California 94305, USA}
\newcommand{\Sussex}{Department of Physics and Astronomy, University of 
Sussex, Falmer, Brighton BN1 9QH, United Kingdom}
\newcommand{\Tennessee}{Department of Physics and Astronomy, 
University of Tennessee, 1408 Circle Drive, Knoxville, Tennessee 37996, USA}
\newcommand{\Texas}{Department of Physics, University of Texas at Austin, 
1 University Station C1600, Austin, Texas 78712, USA}
\newcommand{\Tufts}{Department of Physics and Astonomy, Tufts University, Medford, 
Massachusetts 02155, USA}
\newcommand{\Virginia}{Department of Physics, University of Virginia, 
Charlottesville, Virginia 22904, USA}
\newcommand{\WSU}{Physics Division, Wichita State Univ., 1845 
Fairmout St., Wichita, Kansas 67220, USA}
\newcommand{\WandM}{Department of Physics, College of William \& Mary, 
Williamsburg, Virginia 23187, USA}
\newcommand{\Winona}{Department of Physics, Winona State University, P.O. 
Box 5838, Winona, Minnesota 55987, USA}
\newcommand{\Crookston}{Math, Science and Technology Department, University 
of Minnesota -- Crookston, Crookston, Minnesota 56716, USA}
\newcommand{\deceased}{Deceased.}

\affiliation{\ANL}
\affiliation{\IOP}
\affiliation{\Athens}
\affiliation{\BHU}
\affiliation{\UCLA}
\affiliation{\Caltech}
\affiliation{\Charles}
\affiliation{\Cincinnati}
\affiliation{\Cochin}
\affiliation{\CTU}
\affiliation{\Delhi}
\affiliation{\FNAL}
\affiliation{\UFG}
\affiliation{\Guwahati}
\affiliation{\Harvard}
\affiliation{\Hyderabad}
\affiliation{\IHyderabad}
\affiliation{\Indiana}
\affiliation{\INR}
\affiliation{\Iowa}
\affiliation{\Jammu}
\affiliation{\JINR}
\affiliation{\Lebedev}
\affiliation{\MSU}
\affiliation{\Crookston}
\affiliation{\Duluth}
\affiliation{\Minnesota}
\affiliation{\Oxford}
\affiliation{\Panjab}
\affiliation{\RAL}
\affiliation{\Carolina}
\affiliation{\SDakota}
\affiliation{\SMU}
\affiliation{\Stanford}
\affiliation{\Sussex}
\affiliation{\Tennessee}
\affiliation{\Texas}
\affiliation{\Dallas}
\affiliation{\Tufts}
\affiliation{\Virginia}
\affiliation{\WSU}
\affiliation{\WandM}
\affiliation{\Winona}

\author{P.~Adamson}
\affiliation{\FNAL}

\author{C.~Ader}
\affiliation{\FNAL}

\author{M.~Andrews}
\affiliation{\FNAL}

\author{N.~Anfimov}
\affiliation{\JINR}

\author{I.~Anghel}
\affiliation{\Iowa}
\affiliation{\ANL}

\author{K.~Arms}
\affiliation{\Minnesota}

\author{E.~Arrieta-Diaz}
\affiliation{\SMU}

\author{A.~Aurisano}
\affiliation{\Cincinnati}

\author{D.~S.~Ayres}
\affiliation{\ANL}

\author{C.~Backhouse}
\affiliation{\Caltech}

% ...

\collaboration{The NOvA Collaboration}
\noaffiliation

\date{\today}

\preprint{FERMILAB-PUB-15-262-ND}

\begin{abstract}

We report results from the first search for $\numu{}\rightarrow\nue{}$ transitions by the \nova{} experiment.  In an exposure equivalent to $2.74 \times 10^{20}$ protons-on-target in the upgraded NuMI beam at Fermilab, we observe \lidresult{} events in the Far Detector, compared to a background expectation of \lidbgpredict{} events based on the Near Detector measurement.  A secondary analysis observes \lemresult{} events with a background of \lembgpredict{}.  The \lidsigmaapp{} excess of events observed in the primary analysis disfavors \liddcprange{} in the inverted mass hierarchy at the 90\% C.L.

\end{abstract}

\maketitle


This Letter reports the first \nova{} measurement of the oscillation of muon neutrinos (\numu) into electron neutrinos (\nue) at the first oscillation maximum.   The oscillation probability to first order is proportional to \sinsq{13}, which is well measured by reactor experiments~\cite{ref:reactors}.  Accelerator experiments measuring \numutonue oscillations differ from reactor experiments in that they are sensitive to three physical parameters that are currently unknown or poorly known~\cite{ref:theory}:  \sinsqnotwo{23}, which determines the coupling of \numu to the third neutrino mass state; $\deltacp{}$, which determines the extent to which CP symmetry is violated in the neutrino sector; and the ordering of the neutrino masses, specifically whether the masses of the solar doublet are smaller (normal hierarchy, \nh{}) or larger (inverted hierarchy, \ih{}) than the third neutrino mass. The mass hierarchy may be determined by observing an enhancement (\nh{}) or suppression (\ih{}) of the \numutonue oscillation probability caused by coherent forward scattering of electron neutrinos on electrons in the earth~\cite{ref:msw}.  For a fixed ratio of baseline to neutrino energy, this effect increases with the experiment's baseline.  Previous accelerator measurements of this oscillation mode have been reported by MINOS~\cite{ref:MINOSnue} and T2K~\cite{ref:T2Knue}.  The \nova{} experiment has the longest baseline of any past or present accelerator neutrino oscillation experiment.  

\nova{} uses Fermilab's NuMI neutrino beam, upgraded to allow \unit[700]{kW} maximum power~\cite{ref:NuMI,ref:nova}.  The beam is created by \unit[120]{GeV} protons from the Main Injector striking a \unit[1.2]{m} long graphite target.  Two magnetic horns focus pions and kaons produced in the target.  The focused hadrons decay in a \unit[675]{m} long decay pipe.  The average beam power increased from \unit[250]{kW} to \unit[450]{kW} over the period of data taking.

The \nova{} experiment~\cite{ref:nova} has two detectors located \unit[1]{km} and \unit[810]{km} from the NuMI beam target.  Both are sited \unit[14.6]{mrad} off the central axis of the beam, as measured from the average neutrino production point, where they observe neutrinos mainly in a narrow range of energies between 1 and \unit[3]{GeV}.  These off-axis locations enhance the neutrino flux in the region of the first oscillation maximum and reduce backgrounds, particularly from higher-energy neutral current events.  Simulation predicts that at the position of the Near Detector (ND), the NuMI beam is composed mostly of \numu{} with a 3.8\% \numubar{} component and a 2.1\% (\nue{}+\nuebar{}) component.

The \nova{} detectors are functionally equivalent tracking calorimeters~\cite{ref:detector}, composed of cells of liquid scintillator~\cite{ref:scint} encased in polyvinyl chloride (PVC) extrusions~\cite{ref:extrusions}.  The cross sectional dimension of each cell, including the PVC, is \unit[3.9]{cm} wide by \unit[6.6]{cm} deep.  The extrusions are \unit[15.5]{m} long in the Far Detector (FD) and \unit[3.9]{m} long in the ND.  They are arranged in planes with the long cell dimension alternating between the vertical and horizontal orientations.  The FD (ND) contains 896 (192) planes with a total mass of \unit[14]{\kton{}} (\unit[193]{t}).
To enhance muon containment, the downstream end of the ND has an additional ten layers of 10-cm-thick steel plates interleaved with pairs of one vertical and one horizontal plane of scintillator cells.  In the fiducial region of the detectors, the liquid scintillator comprises 62\% of the detector mass.

The signal from each liquid scintillator cell is read out through a single wavelength-shifting fiber.  The fiber is looped at the far end of the cell, and both near ends of the fiber terminate on the same pixel of a 32-pixel avalanche photodiode (APD)~\cite{ref:apd}.  The APD signal is continuously integrated, shaped, then digitized.  Signals above a preset threshold are sent to a buffer pending a trigger decision~\cite{ref:DAQ}.  All signals within a \unit[550]{$\mu$s} window around the \unit[10]{$\mu$s} NuMI spill are recorded.  Signals from periodic time windows asynchronous to the beam spill are also recorded to collect cosmic rays for calibration.

The data used for this analysis were taken between February 6, 2014 and May 15, 2015.  The FD was under construction until November 2014.
Data collected whenever \unit[4]{\kton{}} or more of contiguous detector mass was operational were used in this analysis.  The effective fiducial mass varied from \unit[2.3]{\kton{}} for \unit[4.0]{\kton{}} of total mass to \unit[10]{\kton{}} for the full \unit[14]{\kton}.  The exposure accumulated was $3.45 \times 10^{20}$ protons on target (\POT{}), equivalent to \unit[$2.74 \times 10^{20}$]{\POT{}} collected in the full \unit[14]{\kton{}} detector.  

The two-detector design of the experiment reduces the reliance on Monte Carlo (MC) simulation, but the simulation still plays an important role in the analysis.  We use {\sc fluka}~\cite{ref:fluka1}, interfaced with a {\sc geant4}~\cite{ref:geant1} geometry using {\sc flugg}~\cite{ref:flugg} to model the interaction of NuMI protons in the \nova{} target, the transport of the products through the target and magnetic field of the horns, and the decay of those products into neutrinos. The interactions of neutrinos in the \nova{} detectors are simulated using {\sc genie}~\cite{ref:genie}, and {\sc geant4} is used to propagate the resulting particles and record energy depositions in the liquid scintillator.  To produce simulated raw signals, or hits, we use experiment-specific simulations to model the capture of scintillation photons in the fibers, light attenuation in the fibers, and the response of the APDs and readout electronics~\cite{ref:sim}.

Raw hits from both data and simulation pass through a series of reconstruction stages~\cite{ref:reco} to produce neutrino interaction candidates. First, collections of hit cells close in space and time are clustered~\cite{ref:michaelthesis,ref:dbscan}, then those clusters are examined to find particle paths~\cite{ref:hough}.  The intersections of the paths are taken as seeds to find the neutrino interaction vertex~\cite{ref:earms}.  The set of cells associated with each of the particle paths emanating from the reconstructed vertex is identified~\cite{ref:fuzzyk, ref:evanthesis}; partial sharing of hits among paths is allowed.  Paths are classified as shower-like based on the transverse energy distribution, and the most energetic shower is designated the primary shower.  Events with a well-defined vertex and reconstructed shower are considered for further analysis.

Raw signals are corrected for light attenuation in the fiber and for cell-to-cell non-uniformity.  Cosmic ray muons that stop in the detector are used as a standard candle for energy calibration~\cite{ref:minoscalib}.  The energy is computed as the sum of the calibrated energy deposited in each cell, using the simulation to correct for the inert material and the energy lost to undetected particles.

The \nova{} FD is on the surface, beneath a modest overburden which blocks most of the electromagnetic component of cosmic ray secondaries.  To further reject backgrounds from these events, we require that selected events are in a \unit[12]{$\mu$s} time window around the beam spill.
Additionally, showers must be well separated from the edges of the detector~\cite{ref:tianthesis}.  Restricting the distance of the primary shower from the detector edges also removes events on the periphery of the detector.
The containment requirements are more stringent at the top and back of the detector where most of the cosmic background events enter the volume.  Additionally, steep events that likely originate from cosmic rays are rejected.  These selection criteria were determined using a large sample of calibration data.  To measure the cosmic background, the rejection criteria are applied to the independent data set collected during the \unit[550]{$\mu$s} around the beam spill, excluding a \unit[30]{$\mu$s} window centered on the spill.  This sample reproduces the detector configuration and data quality conditions of the data in the beam spill.

To observe  \numutonue{} oscillations, electron neutrino charged-current interactions (\nueCC{}) must be identified in the FD.  These interactions are characterized by an electron cascade, along with other potential activity produced by the breakup of the recoil nucleus.  The size of the electromagnetic cascade is characterized by the detector Moli\`ere radius of $\sim$3 cell widths and radiation length of $\sim$6 planes.  The combination of the beam energy spectrum and the energy-dependent nature of the oscillation means the maximal \nue{} signal appears around \unit[2]{GeV}.

The interactions of the beam \nue{} component are a background to the analysis.
Neutral-current (NC) and \numuCC{} interactions are also backgrounds to this analysis,  particularly when the hadronic recoil system contains a \piz{}. The \numuCC{} are a relatively small background in the FD as they are suppressed by oscillations.  Even less significant are \nutauCC interactions from $\numu{}\rightarrow\nutau{}$ oscillations and \numubar{} from the beam.  NC events and cosmic ray induced events populate the low energy range, while beam \nueCC{} events tend to be at higher energies.  Therefore, we select neutrino interaction candidates with a total calorimetric energy of 1.3 to \unit[2.7]{GeV}.
Additional requirements on the number of occupied cells in the event and the length of the longest particle path suppress clear non-\nueCC{} interactions.  

\begin{figure}[tb!]
  \includegraphics[width=\linewidth]{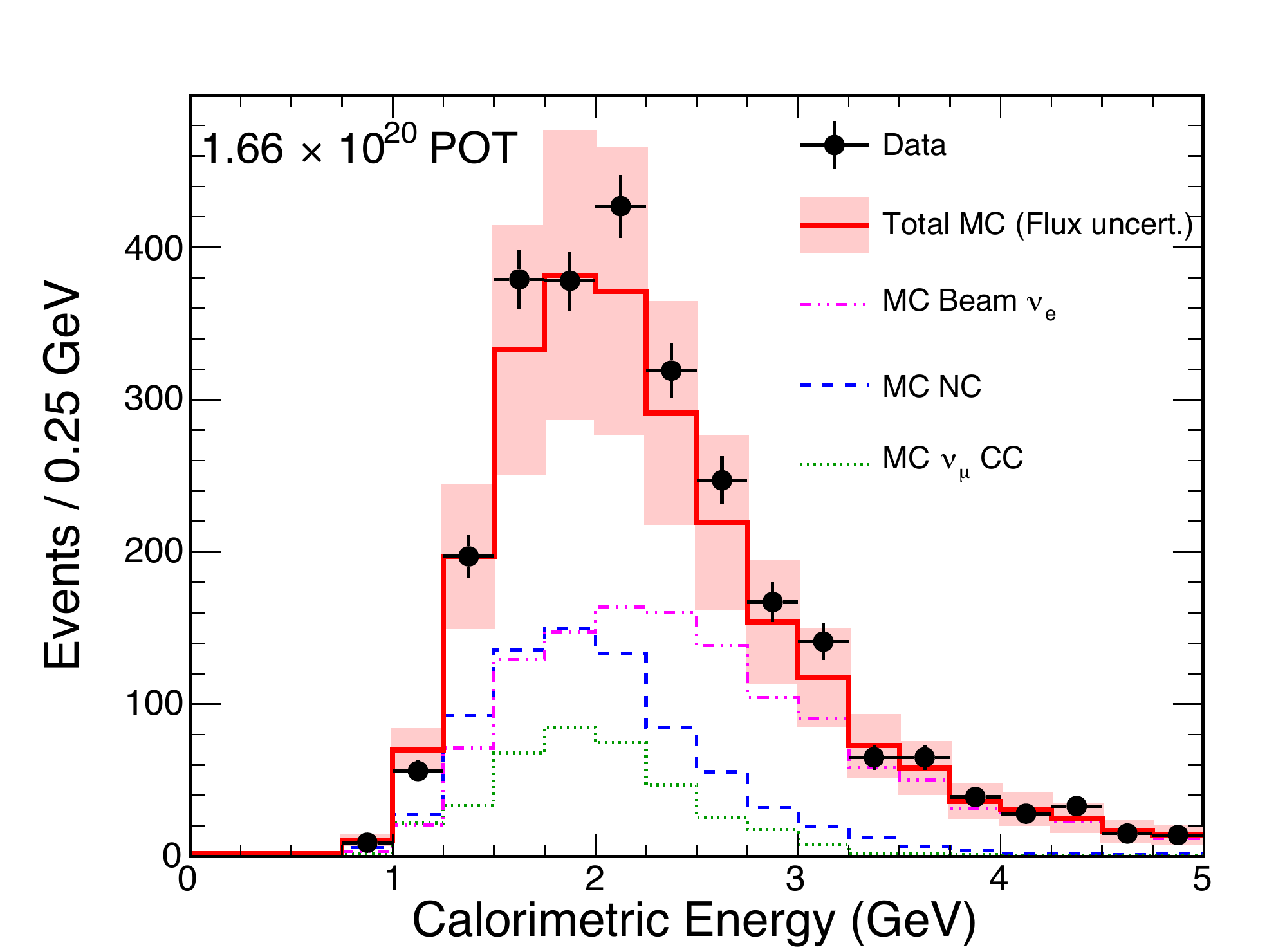}
  \caption{Reconstructed energy distribution for events selected with the primary selector in the ND data and MC simulation.  Events selected with the secondary selector show similar agreement between data and simulation.}
  \label{fig:nd}
\end{figure}

To further enhance the \nueCC{} sample purity, more sophisticated algorithms are necessary. A first method, a likelihood-based selector (LID), compares the longitudinal and transverse energy deposition in the primary shower to template histograms for various simulated particles~\cite{ref:evanthesis, ref:kanikathesis, ref:jbiantext}.  The likelihood differences among different particle hypotheses and other topological variables are used as input to an artificial neural network to construct the primary classifier.  The energy range of events selected with this primary method is further restricted to 1.5 to \unit[2.7]{GeV} to remove additional backgrounds from cosmic radiation. 

A second selection method, Library Event Matching (LEM), compares an input event from either data or simulation to a large and independent library of simulated events~\cite{ref:LEM}. The properties of the library events that are most similar to the input event provide information about the most likely identity of the neutrino interaction.  This and additional identifying information from the best matches in the library is fed into an ensemble decision tree that gives the final classifier for this technique.

Both selectors achieve similar signal efficiency and background rejection of simulated events.  The LID selection method achieves a signal efficiency of 34\% relative to the event sample meeting the containment criteria, while the LEM selection is 35\% efficient.  Simulations predict a 62\% overlap in the signal events chosen.  Both classifiers reject 99\% of beam backgrounds.  Each of the selection techniques achieves a rejection better than 1 in $10^{8}$ for cosmic induced backgrounds.  The more traditional LID selection was chosen as the primary selection technique, but it was agreed that results from LEM would also be presented.  This choice and all other analysis techniques were finalized before inspecting the FD beam data.

Similar selection criteria are applied to the ND sample, where all events are background events.  Energy cuts are not applied in the ND so the full spectrum can be inspected.  Figure~\ref{fig:nd} shows the reconstructed energy spectrum of the events passing the primary selector in the ND data, compared to the simulation, which is normalized to the same exposure.  About 7\% more background events are selected in the data relative to the simulation.  

The FD beam-induced background is predicted by scaling the number of events selected in the FD simulation by the observed ND ratio of data to simulation in each bin of reconstructed energy.  Each background component is scaled by the same factor.  The FD simulated events are weighted by the three flavor oscillation probability~\cite{ref:barger}.  The small number of expected \nutau{} background events is taken directly from the FD simulation.  The predicted background from cosmic radiation and the beam, broken down by component, is given in Table~\ref{tab:bgpred} for both selection techniques~\cite{ref:oscpars}.  

\begin{table}[htb!]
\centering
\begin{tabular}{r|c c c c c|c}
      &~Beam\,\nue~&~NC~&~\numuCC~&~\nutauCC~&~Cosmic~& Total Bkg.\\
  \hline
  LID &   0.50    &  0.37  &   0.05  &   0.02   &  0.06  &   0.99\\
  LEM &   0.50    &  0.43  &   0.07  &   0.02   &  0.06  &   1.07\\
\end{tabular}
\caption{Predicted number of background events for each of the event selection techniques.}
\label{tab:bgpred}
\end{table}

The number of signal events expected from \nue{} appearance is also derived from the ND data.  The energy spectrum of \numuCC{}-selected events~\cite{ref:michaelthesis,ref:susanthesis,ref:nickthesis} in the ND is compared to the simulation and the discrepancy between the two is interpreted as an inexact modeling of the underlying true energy spectrum.  The FD simulated energy spectrum for \nue{} events is adjusted to account for the discrepancy, increasing the predicted signal by 1\%.  With the oscillation parameters given in~\cite{ref:oscpars} 5.2 (5.4) signal events from $\numu{}\rightarrow\nue{}$ are expected to pass the LID (LEM) selection criteria.

\begin{table}[tb!]
  \begin{center}
    \begin{tabular}{r|cc}
      & Signal (\%)&Bkg. (\%)\\
      \hline
      Calibration & 7.6&4.4\\
      Neutrino interaction &14.0&3.7\\
      Scintillator saturation &7.2& 5.1 \\
      Normalization& 1.2 & 1.2 \\
      Neutrino flux & 1.1& 3.2 \\
      ND bkg. composition  &--& 5.4 \\
      Other&0.6&3.9\\
      \hline
      Total & 17.6 & 10.8\\
    \end{tabular}
  \end{center}
  \caption{Systematic uncertainty on the background and signal prediction for events selected by the primary selector in the FD.  The last row corresponds to the quadrature sum.}
  \label{tab:systs}
\end{table}

While the two-detector technique mitigates the impact of many sources of systematic uncertainty, some residual uncertainties remain.  These uncertainties are evaluated by modifying the simulation to account for the different sources of uncertainty, then generating new simulated events.  Background and signal predictions are made using the modified sample; the change in the number of events predicted compared to the nominal simulation is used to quantify the size of each effect.  The effects considered are tabulated in Table~\ref{tab:systs}.  

Dominant sources of uncertainty in the signal prediction arise from uncertainties in the modeling of neutrino-nucleus interactions, including a 40\% uncertainty on the value of the axial-vector mass of \unit[0.99]{GeV/c$^2$} used in the quasielastic scattering model~\cite{ref:genie, ref:jithesis}.  The allowed variation in this effective parameter encompasses recent measurements~\cite{ref:K2K, ref:MiniBooNE, ref:minosqema, ref:t2kqema} and is a proxy for possible multinucleon processes not included in the interaction model~\cite{ref:Martini, ref:Valencia, ref:MINERvA, ref:minerva2p2h}.  Dominant sources of uncertainty affecting the background prediction include a 5\% uncertainty on both the absolute energy calibration and the inter-detector energy calibration, uncertainty in the modeling of scintillator saturation by highly ionizing particles~\cite{ref:birks}, and modeling of the neutrino flux.  The error incurred by scaling each background component by the same amount, instead of employing a data-driven decomposition of the background components, is estimated by individually scaling each background component to account for the entire difference between data and simulation.  

An overall normalization uncertainty on both signal and background levels in the FD comes from a survey of the mass of the materials used in the ND relative to the FD, combined with uncertainty in the measurement of \POT{} delivered as well as a small difference between data and simulation in the efficiency for reconstructing events.  Other considerations include possible biases arising from different containment criteria in the ND relative to the FD, imperfect removal of uncontained vertex events, and limited statistics in both the simulation and the ND data set.  Adding all the effects in quadrature gives a 17.6\% (15.0\%) systematic uncertainty on the signal prediction and a 10.8\% (13.4\%) systematic uncertainty on the background prediction for the primary (secondary) selection technique.

\begin{figure}[tb!]
  \includegraphics[width=\linewidth]{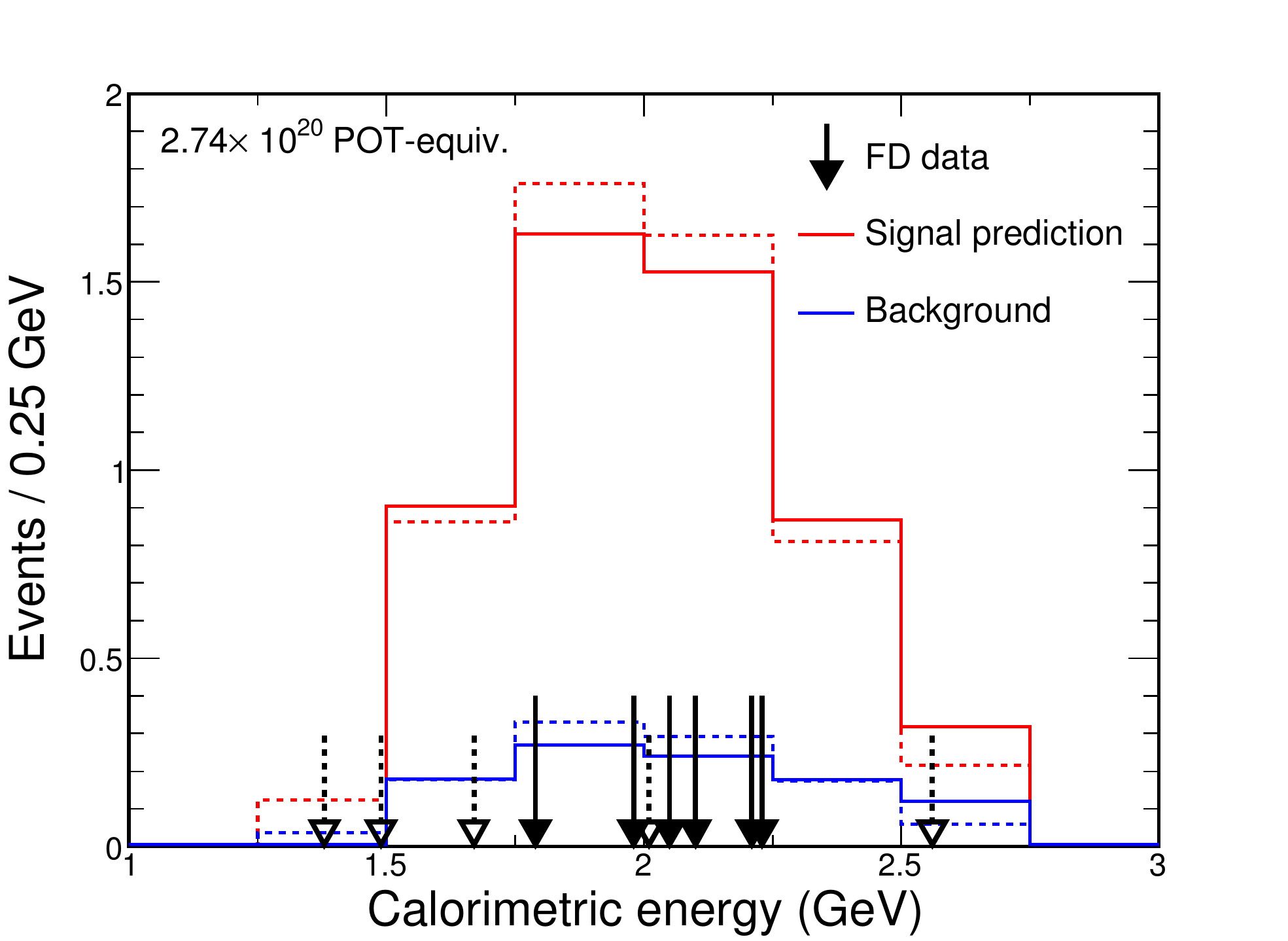}
  \caption{Reconstructed energy distribution of events selected in the FD.  Solid (dotted) histograms show the prediction for the primary (secondary) selector.  Arrows indicate where the data lie.  Solid arrows show events from the primary selector, while dotted arrows show the additional events from the secondary.}
  \label{fig:fd}
\end{figure}

Upon examining the FD data, \lidresult{} events were observed, compared to the background prediction of \lidbgpredict{}.  The observation corresponds to a \lidsigmaapp{} excess over the background prediction.  With the secondary event selection, we observe \lemresult{} events, a \lemsigmaapp{} excess over the background prediction of \lembgpredict{}.  All the events selected by the primary selector are in the sample selected by the secondary.  Using the trinomial probability distribution and the number of simulated events that overlap between the selectors or are selected by each exclusively, we compute a 7.8\% probability of observing our particular overlap configuration or a less likely configuration.  Figure~\ref{fig:fd} shows the energy distribution in the FD for events selected by either selection technique compared to the predicted spectrum with oscillation parameters as given in~\cite{ref:oscpars}.

\begin{figure}[tb!]
  \begin{centering}
    \includegraphics[trim=0 0 31 0,clip=true,width=.509\linewidth]{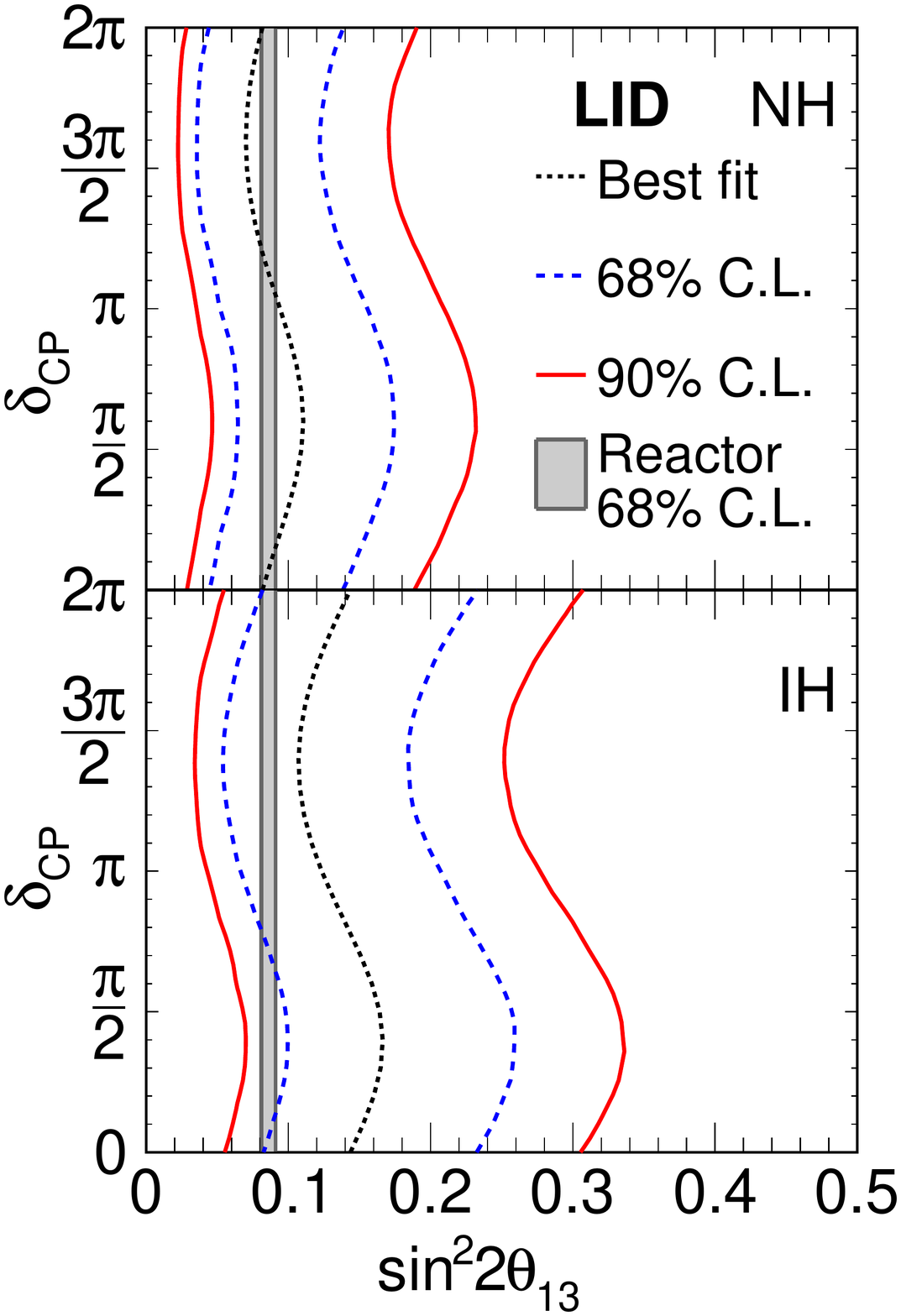}
    \includegraphics[trim=69 0 0 0,clip=true,width=.471\linewidth]{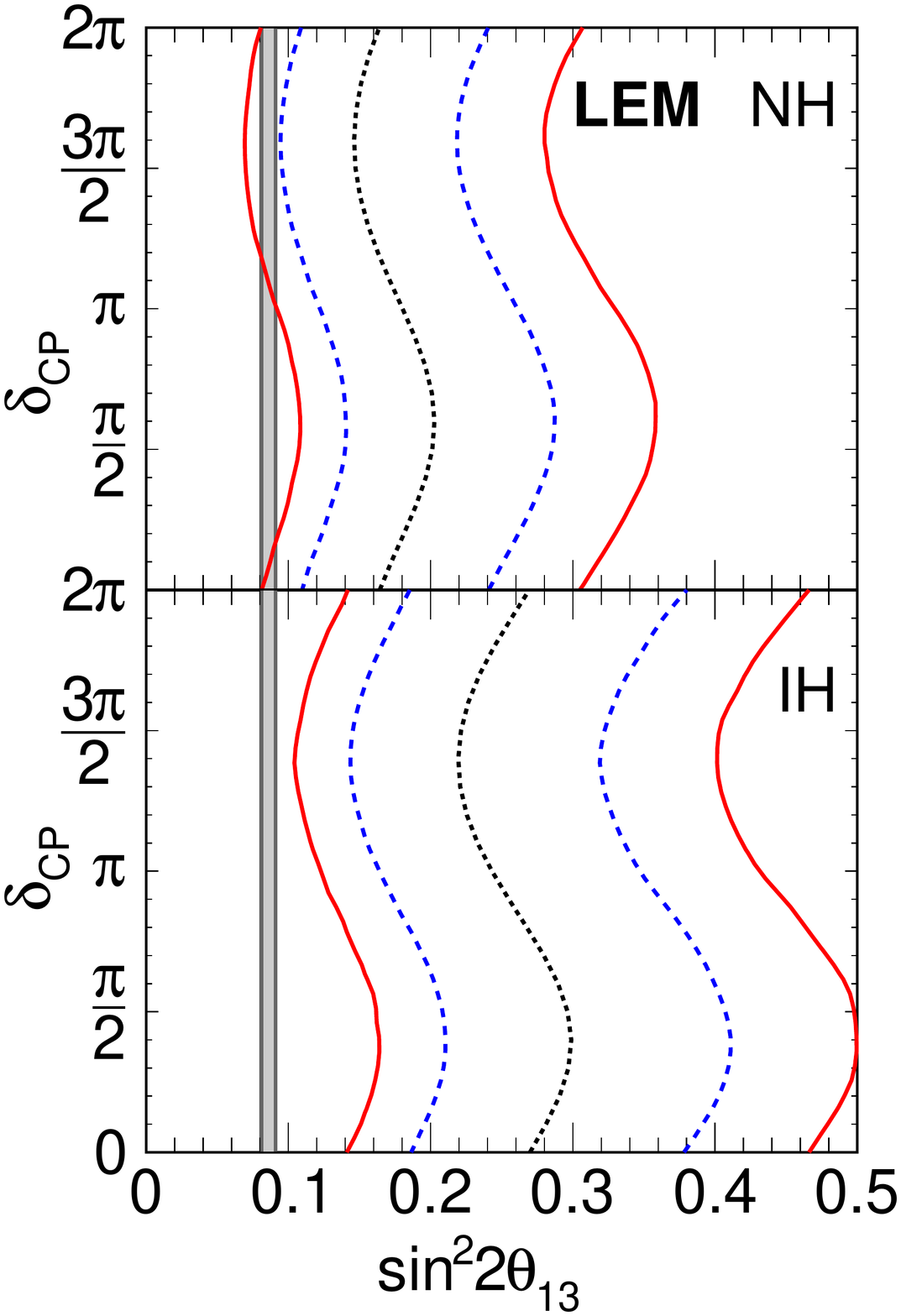}\\
  \end{centering}
  \caption{Allowed values of  $\deltacp{}$ vs $\sinsq{13}$.
Top (bottom) plots show the \nh{} (\ih{}).  Left (right) plots show results for the primary (secondary) selector.  Both have $\sin^2\theta_{23}$ fixed at 0.5.}
  \label{fig:contours}
\end{figure}

The likelihood for a Poisson distributed variable is used to compare the observed number of events to that predicted for a particular set of oscillation parameters.  Figure~\ref{fig:contours} shows the values of $\deltacp{}$ and $\sin^2{2\theta_{13}}$ consistent with the observed number of events in the data for each of the selectors.  Following the procedure of Feldman and Cousins~\cite{ref:fc}, we determine confidence intervals by inspecting the range of likelihood ratios observed in pseudo-experiments.  Uncertainties in signal and background predictions, in the solar oscillation parameters, and in the atmospheric mass splitting~\cite{ref:PDG} are included in the generation of these pseudo-experiments, while $\sin^2\theta_{23}$ is fixed at 0.5.  The data selected by the primary selector are compatible with three-flavor oscillations at the reactor value of $\theta_{13}$.  The number of events selected by the secondary selector favors a higher value of $\sinsq{13}$ for $\sin^2\theta_{23}$ fixed at 0.5, or alternatively a higher value of $\sin^2\theta_{23}$ for $\sinsq{13}$ constrained to the reactor measurement.

\begin{figure}[tb!]
  \includegraphics[width=\linewidth]{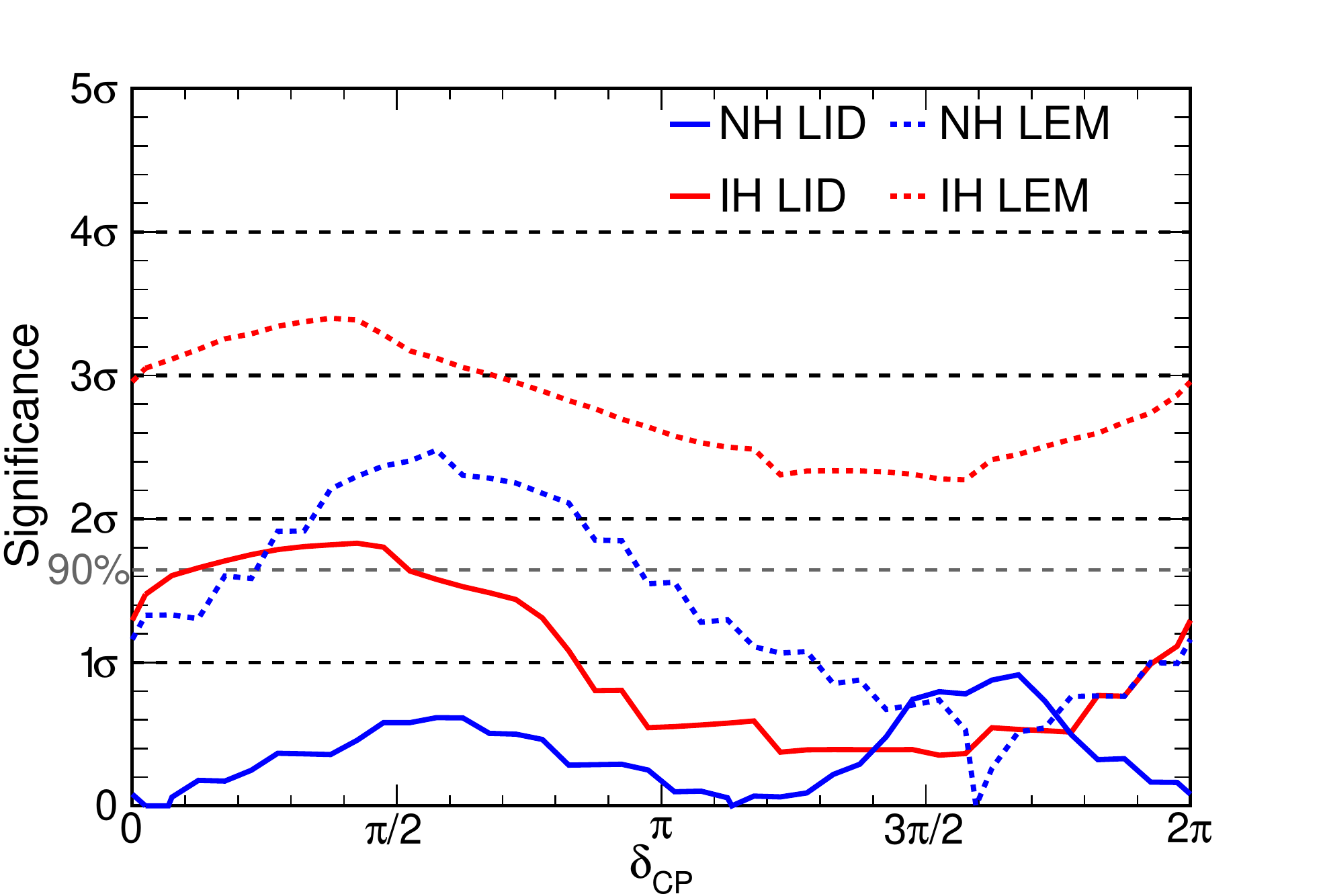}
  \caption{Significance of the difference between the selected and the predicted number of events as a function of $\deltacp{}$ and the hierarchy.  The primary (secondary) selection technique is shown with solid (dotted) lines.}
  \label{fig:chisq_scans}
\end{figure}

Figure~\ref{fig:chisq_scans} shows the compatibility between the observation and the number of events expected as a function of the mass hierarchy and $\deltacp{}$ if we additionally assume the reactor constraint of $\sin^22\theta_{13}=0.086\pm 0.005$~\cite{ref:reactors}.  The maximal mixing constraint is also removed, and uncertainty in  $\sin^2\theta_{23}$ is included in the generation of the pseudo-experiments~\cite{ref:PDG}.  For each value of $\deltacp{}$ and choice of hierarchy we compute the likelihood ratio to the best fit parameters and show the fraction of pseudo-experiments which have a larger or equal likelihood ratio, converted into a significance.  The discontinuities are due to the discrete set of possible event counts.  The range of \liddcprange{} in the \ih{} is disfavored at the 90\% C.L.  The number of events selected by the secondary analysis is larger than the number of events expected given the range of oscillation parameters favored in global fits~\cite{ref:globalfits}, but 13\% of pseudo-experiments generated at the NOvA best fit find at least as many events as observed in the data.  With the secondary selector all values of $\deltacp{}$ in the \ih{} are disfavored at greater than 90\% C.L. The range of \lemdcprangenh{} in the \nh{} is disfavored at the 90\% C.L. 

In conclusion, with an exposure of $2.74\times 10^{20}$ \POT{}, \nova{} observes \lidresult{} \nue{}-like events in the FD, with a background prediction of \lidbgpredict{}.  The \lidsigmaapp{} excess of events above background disfavors \liddcprange{} in the inverted mass hierarchy at the 90\% C.L.

This work was supported by the US Department of Energy; the US National Science Foundation; the Department of Science and Technology, India; the European Research Council; the MSMT CR, Czech Republic; the RAS, RMES, and RFBR, Russia; CNPq and FAPEG, Brazil; and the State and University of Minnesota. We are grateful for the contributions of the staffs at the University of Minnesota module assembly facility and Ash River Laboratory, Argonne National Laboratory, and Fermilab. Fermilab is operated by Fermi Research Alliance, LLC under Contract No.~De-AC02-07CH11359 with the US DOE.


\begin{thebibliography}{99}
  \bibitem{ref:reactors} F.~P.~An \etal, Phys. Rev. Lett. {\bf 115}, 111802 (2015);  S.~H.~Seo arXiv:1410.7987; Y.~Abe \etal, JHEP {\bf 1410}  86 (2014).  For calculations in this Letter, we use the weighted average, $\sinsq{13} = 0.086 \pm 0.005$.
  \bibitem{ref:theory} A.~Cervera \etal, Nucl.\ Phys. {\bf 579}, 17 (2000); A.~Cervera \etal, Nucl.\ Phys.\ B {\bf 593}, 731 (2001).
  \bibitem{ref:msw} L.~Wolfenstein, Phys. Rev. D {\bf 17}, 2369 (1978); S.~P.~Mikheyev and A.~Yu.~Smirnov, Sov. J. Nucl. Phys. {\bf 42}, 913 (1985).
  \bibitem{ref:MINOSnue} P.~Adamson \etal, Phys. Rev. Lett. { \bf 110}, 171801 (2013).
  \bibitem{ref:T2Knue}  K.~Abe \etal, Phys. Rev. Lett. {\bf  112}, 061802 (2014).
  \bibitem{ref:NuMI} P.~Adamson \etal, Nucl. Instrum. Meth. A {\bf 806}, 279 (2016); NuMI Technical Design Handbook, FERMILAB-DESIGN-1998-01.

  \bibitem{ref:nova} \nova{} Technical Design Report, FERMILAB-DESIGN-2007-01.
    
  \bibitem{ref:detector} S.~Magill, J. Phys.: Conf. Ser. {\bf  404}, 012035 (2012); P.~Border, \etal, Nucl. Instrum. Meth. A {\bf 463}, 194 (2001).

  \bibitem{ref:scint} S.~Mufson \etal, Nucl. Instrum. Meth. A {\bf 799}, 1 (2015).

  \bibitem{ref:extrusions} R.~L.~Talaga, \etal, FERMILAB-PUB-15-049-ND-PPD.

  \bibitem{ref:apd} The \nova{} APD is a custom variant of the Hamamatsu S8550.  \url{http://www.hamamatsu.com/us/en/product/alpha/S/4112/S8550-02/index.html}.
  \bibitem{ref:DAQ} J.~Z\'{a}le\u{s}\'{a}k \etal, J. Phys.: Conf. Ser. {\bf  513}, 012041 (2014).
  \bibitem{ref:fluka1} T.~Bohlen \etal, Nucl. Data Sheets {\bf 120}, 211 (2014); A.~Ferrari \etal, Tech. Rep. CERN-2005-010 (2005).

  \bibitem{ref:geant1} S.~Agostinelli \etal, Nucl. Instrum. Meth. A {\bf 506}, 250 (2003); J.~Allison \etal, IEEE Trans. Nucl. Sci. {\bf 53}, 270 (2006).

  \bibitem{ref:flugg}  M.~Campanella \etal, Tech. Rep. CERN-ATL-SOFT-99-004 (1999).

  \bibitem{ref:genie} C.~Andreopoulos \etal, Nucl. Instrum. Meth. A {\bf 614}, 87 (2010);  C.~Andreopoulos \etal, arXiv:1510.05494.

  \bibitem{ref:sim} A.~Aurisano \etal, J. Phys.: Conf. Ser. {\bf 664}, 072002 (2016).

  \bibitem{ref:reco} M. Baird \etal,  J. Phys.: Conf. Ser. {\bf 664} 072035 (2016).
  \bibitem{ref:michaelthesis} M.~Baird, Ph.D. Thesis, Indiana University (2015).

  \bibitem{ref:dbscan} M.~Ester \etal, Proc. of 2nd International Conference on Knowledge Discovery and Knowledge Engineering and Knowledge Management, 226 (1996).
  \bibitem{ref:hough} L.~Fernandes and M.~Oliveira, Patt. Rec. {\bf 41}, 299 (2008).
  \bibitem{ref:earms} M.~Gyulassy and M.~Harlander, Comput. Phys. Commun. {\bf 66}, 31 (1991); M.~Ohlsson, and C.~Peterson, Comput. Phys. Commun. {\bf 71}, 77 (1992); M.~Ohlsson, Comput. Phys. Commun. {\bf 77}, 19 (1993); R~Fr\"{u}hwirth and A.~Strandlie, Comput. Phys. Commun. {\bf 120}, 197 (1999).
  \bibitem{ref:fuzzyk} R.~Krishnapuram and J.~M.~Keller, IEEE Trans. Fuzzy Syst. {\bf 1}, 98 (1993); M.~S.~Yang, and K.~L.~Wu, Pattern Recognition {\bf 39}, 5 (2006).
  \bibitem{ref:evanthesis} E.~Niner, Ph.D. Thesis, Indiana University (2015).
  \bibitem{ref:minoscalib} D.~G.~Michael \etal, Nucl. Instrum. Meth. A {\bf 596}, 190 (2008).
  \bibitem{ref:tianthesis} T.~Xin, Ph.D. Thesis, Iowa State University (2016).
  \bibitem{ref:kanikathesis} K.~Sachdev, Ph.D. Thesis, University of Minnesota (2015).
    \bibitem{ref:jbiantext} J.~Bian, arXiv:1510.05708.
    \bibitem{ref:LEM} C.~Backhouse and R.~B.~Patterson, Nucl. Instrum. Meth. A {\bf 778}, 31 (2015).
      \bibitem{ref:barger} V.~Barger \etal, Phys. Rev. D {\bf 22}, 2718 (1980); J.~Kopp, Int. J. Mod. Phys. C {\bf 19}, 523 (2008).
  \bibitem{ref:oscpars} Specifically, $\sinsqnotwo{23}=0.5$, $\dmsq{32}=+2.37\times10^{-3}{\rm eV^{2}}$, $\sinsq{12}=0.846$, $\dmsq{21}=7.53\times10^{-5}{\rm eV^2}$, $\sinsq{13}=0.086$, and $\deltacp{}=0$.  Backgrounds vary at the few percent level for different choices of oscillation parameters.

  \bibitem{ref:susanthesis} S.~Lein, Ph.D. Thesis, University of Minnesota (2015).
  \bibitem{ref:nickthesis} N.~Raddatz, Ph.D. Thesis, University of Minnesota (2016).
  \bibitem{ref:jithesis} J.~Liu, Ph.D. Thesis, College of William and Mary (2016).
  \bibitem{ref:K2K} R.~Gran \etal, Phys. Rev. D {\bf 74}, 052002 (2006).
  \bibitem{ref:MiniBooNE} A.~A.~Aguilar-Arevalo \etal, Phys. Rev. D {\bf 81}, 092005 (2010).
  \bibitem{ref:minosqema} P.~Adamson \etal, Phys. Rev. D {\bf 91}, 012005 (2015).
  \bibitem{ref:t2kqema} K.~Abe \etal,  Phys. Rev. D {\bf 91}, 112002 (2015). 
  \bibitem{ref:Martini} M.~Martini \etal, Phys. Rev. C {\bf 80}, 065501 (2009);  Phys. Rev. C {\bf 81}, 045502 (2010). 
  \bibitem{ref:Valencia} R.~Gran \etal, Phys. Rev. D {\bf 88}, 113007 (2013). 
  \bibitem{ref:MINERvA} G.~A.~Fiorentini \etal, Phys. Rev. Lett. {\bf 111}, 022502 (2013).
  \bibitem{ref:minerva2p2h} P.~A.~Rodrigues \etal, arXiv:1511.05944.
    \bibitem{ref:birks}J.~B.~Birks, Proc. Phys. Soc., {\bf A64}, 874 (1951).
  \bibitem{ref:fc} G.~J.~Feldman and R.~D.~Cousins, Phys. Rev. D {\bf 57}, 3873 (1998).
  \bibitem{ref:PDG} K.~A.~Olive \etal (Particle Data Group), Chin. Phys. C {\bf 38}, 090001 (2014) and 2015 update.
  \bibitem{ref:globalfits} M.~C.~Gonzalez-Garcia \etal, J. High Energy Phys. {\bf 1411},  052 (2014); F.~Capozzi \etal, Phys. Rev. D {\bf 89}, 093018 (2014).
\end{thebibliography}
\end{document}